\documentclass[a4paper,11pt]{article}
\pdfoutput=1 

\usepackage{jcappub} 

\usepackage[T1]{fontenc} 

\usepackage{graphicx}
\usepackage[margin=1in]{geometry}
\usepackage[utf8]{inputenc}
\usepackage[section]{placeins}
\usepackage{color}
\usepackage{hyperref}
\usepackage{eqnarray}
\usepackage{tikz}
\usepackage{subfig}
\usepackage{tensor}
\usepackage{amsmath,amsthm,amssymb}
\usepackage[ruled,vlined]{algorithm2e}
\usepackage{multicol}
\usepackage{float}
\usepackage{blindtext}
\usepackage{makecell}
\title{Constraining Generalized Chaplygin Gas in Non-Minimally Coupled $f(Q)$ Cosmology using Quasars and $H(z)$ Data}
\author[a,1]{Nakul Aggarwal\note{Equal contribution}}
\author[a,1]{Ali Pourmand}
\author[b,2]{Fatimah Shojai\note{Corresponding Author}}
\author[c]{Harish Parthasarathy}
\affiliation[a]{Department of Physics, University of Alberta, Edmonton, T6G 2E7, Alberta, Canada}
\affiliation[b]{Department of Physics, University of Tehran,
Tehran, Iran.}
\affiliation[c]{Department of Electronics and Communication Engineering, Netaji Subhas University of Technology, New Delhi, India}
\date{\vspace{-5ex}}
\emailAdd{naggarw@ualberta.ca}
\emailAdd{pourmand@ualberta.ca}
\emailAdd{fshojai@ut.ac.ir}
\emailAdd{harishp@nsit.ac.in}

\abstract{In the current framework of Einstein's equations in general relativity (GR), gravity is described by the spacetime curvature. However, there are other descriptions where the origin of gravity can be understood through torsion and non-metricity $Q$. In this work, we discuss a modified theory of gravity namely $f(Q)$ gravity, which considers a non-linear extension of $Q$. In particular, we study the case where it is non-minimally coupled to matter. Motivated by the recent success of Chaplygin gas models in the explanation of dark energy, we assume a pressureless baryonic matter and a generalized Chaplygin gas as the background fluid. We constrain the proposed model using two different datasets: one for Hubble measurements and the other for quasars (which we calibrated) with Markov-Chain Monte Carlo (MCMC) methods. We employ kinematic tools such as deceleration and jerk parameters to determine deviations of the proposed model from $\Lambda$CDM. We establish that the transition redshift $z_T$ in the deceleration parameter $q$ is $0.607$ and $0.204$ with the two datasets respectively, therefore describing the universe's acceleration.}


\begin{document}
\maketitle
\flushbottom

\section{Introduction}
\label{sec:intro}
Based on the observations \cite{riess1998observational,perlmutter1999measurements,eisenstein2005detection,percival2010baryon}, it is believed that the universe is accelerating. One of the most plausible theories for this acceleration is the existence of dark energy with negative pressure. Among all the possible candidates for dark energy, $\Lambda$CDM cosmology has emerged out to be a relatively successful model, as it has been able to explain many cosmological phenomena like the formation of large-scale structures and accurately describes the type Ia Supernovae (SNe Ia) observations. However, it suffers from major problems, namely the fine-tuning and the cosmic coincidence \cite{copeland2006dynamics}. To alleviate these problems, many new models of modified gravity have been proposed. In this regard, $f(R)$ gravity \cite{de2010f}, where the Lagrangian is a function of the scalar curvature $R$, has explained the accelerated expansion of the universe without the need for dark energy or dark matter. However, in the current General Relativity (GR) framework, spacetime is torsionless. But, if one relaxes the need for a metric-compatible connection, new modified theories can be constructed.

Torsion and non-metricity, besides curvature, can also represent the affine properties of a manifold. This gave rise to a theory called the "Teleparallel Equivalent to GR" (TEGR) \cite{buchdahl1970non, Aldrovandi2012TeleparallelGA}, in which the underlying gravity is described by torsion $T$. The corresponding action $S=\int d^4x\sqrt{-g}T$ leads to a TEGR theory with the resulting field equations of order two as compared to the fourth order field equations from $f(R)$ theories. In \cite{jimenez2018coincident}, a modification of TEGR namely "Symmetric Teleparallel Equivalent to GR" (STEGR) was proposed in which the underlying gravitational interaction is described by the non-metricity $Q$ with no torsion and curvature. In a non-Riemannian geometry, $Q$ measures the change of vector length when it is being parallel transported. $f(Q)$ gravity \cite{jimenez2020cosmology} is the extension of the STEGR theory in which the action is described by $S=\int d^4x\sqrt{-g}f(Q)$, where $f(Q)$ is an arbitrary function of $Q$.

The authors in \cite{lazkoz2019observational} have constrained various functional forms of $f(Q)$ using the cosmological observations from type Ia supernovae, baryon acoustic oscillations (BAO), and quasars. $\sqrt{Q}$ corrections were considered in \cite{atayde2021can}, where it was established that this model gave a better fit to observational data in comparison to $\Lambda$CDM through evaluation of the deviance information criterion. For more recent detailed studies on $f(Q)$ gravity, we refer the reader to \cite{barros2020testing,dimakis2021quantum,khyllep2021cosmological,khyllep2022cosmology,mandal2020cosmography,mandal2020energy}. However, all these models assumed minimal coupling of $f(Q)$ gravity with the matter. Interestingly, non-minimal coupling to matter was recently studied in \cite{harko2018coupling}, where the underlying action is $S=\int d^4x\sqrt{-g}\{f_1(Q)/2+f_2(Q)\mathcal{L}_m\}$. Here, $f_1(Q)$ and $f_2(Q)$ are two arbitrary functions of $Q$ and $\mathcal{L}_m$ is the matter Lagrangian. Assuming a perfect fluid matter distribution, observational constraints on the equation of state parameters in power-law non-minimally coupled $f(Q)$ cosmology were obtained in \cite{mandal2021constraint} and the authors showed that $f(Q)$ gravity displays quintessence behavior.

In contrast to the universe with a perfect fluid, many authors have used the Chaplygin gas (CG) model as a possible candidate for dark energy. Its corresponding equation of state is $p=-A/\rho$, where $p$ and $\rho$ are pressure and energy density respectively and $A$ is a constant. The underlying reason for the selection of Chaplygin gas is that it gives a unified picture of dark matter and dark energy. In the early universe, it behaves like a pressure-less dark matter and like the cosmological constant in the late-time universe. However, it has complications with the cosmological power spectrum \cite{sandvik2004end}. To alleviate this problem, the authors in \cite{kamenshchik2001alternative} introduced a phenomenological extension of CG which interpolates between dust and dark-energy dominated era, namely the "Generalized Chaplygin Gas (GCG)" \cite{bento2002generalized}. GCG is a perfect fluid, however, with a polytropic equation of state $p=-A/\rho^n$, where $0<n\le1$. 

In this work, we investigate non-minimally coupled $f(Q)$ gravity with quadratic corrections \cite{lu2019cosmology} with a pressureless baryonic matter and GCG as the background fluid. We test the accuracy of this model by using two different datasets; one is from a variety of direct measurements of the Hubble parameter $H(z)$ at different redshifts $z$, and the other is from measurements of the various properties of quasars. 

Quasars are active galactic nuclei with very high persistent luminosity. They are among the furthest (and oldest) objects that can be observed in the universe. 
Their potential to investigate cosmological models gains more importance once one considers that a notable number of relatively high redshifts quasars have recently been discovered, thanks to projects such as the Sloan Digital Sky Survey (SDSS) \cite{2020sdss}.

One method to use quasars in order to test the validity of various cosmological models is to use them as "standard candles"; in the same way that Cepheid Variables \cite{1929hubble}, and more recently type Ia Supernovae \cite{riess1998observational} have been  used. This has become possible thanks to the work of \cite{Lusso_2020}, who have developed a technique to deduce the distance moduli of quasars. This technique is based on the observed relation between the X-ray and Ultraviolet luminosity of quasars \cite{tananbaum1979}.

This work is organized as follows: In Sec. \ref{sec:model}, we introduce
the $f(Q)$ gravity coupled non-minimally to the matter fields and determine the modified Einstein's equations. In Sec. \ref{sec:flatflrw}, we find the field equations against the backdrop of a spatially flat Friedmann-Lemaitre-Robertson-Walker (FLRW) space-time universe. We assume pressureless baryonic matter and generalized Chaplygin gas for background fluid. In Sec. \ref{sec:data}, we describe the two datasets, $H(z)$ and quasars, and the MCMC method used to constrain the models with these datasets. 
In Sec. \ref{sec:results} and Sec. \ref{sec:diagnostics}, we show our parameter constraining results and evaluate kinematic diagnostics such as deceleration and jerk parameters to analyze deviations from $\Lambda$CDM. Finally in Sec. \ref{sec:conclusions}, we summarize our main findings and their implications, compare them with previous studies and discuss future perspectives.
\section{$f(Q)$ Model}
\label{sec:model}
Similar to non-minimally coupled $f(R)$ gravity \cite{thakur2011non}, the action for the non-minimally coupled $f(Q)$ cosmology \cite{jimenez2018coincident,mandal2021constraint} is given as 
\begin{equation}
\label{eq:Action}
    S=\int d^4x \sqrt{-g}\left[\frac{1}{2\kappa^2}f_1(Q)+f_2(Q)\mathcal{L}_m \right]
\end{equation}
where $f_1(Q)$ and $f_2(Q)$ are functions of non-metricity $Q$, $\mathcal{L}_m$ is the matter Lagrangian density, $g\equiv \text{det}(g_{\mu\nu})$, $g_{\mu\nu}$ is the underlying metric and $\kappa^2=8\pi \mathcal{G}$ where $\mathcal{G}$ is the gravitational constant. Here, we set $\kappa^2=1$. If $f_2(Q)=1$ and $f_1(Q)=Q$ , we retrieve the standard well-studied $f(Q)$ cosmology \cite{jimenez2018coincident}. 
The non-metricity Q is given as
\begin{equation}
\label{eqn:non-metricity}
    Q=g^{\mu\nu}(L\indices{^\lambda_{\sigma\lambda}}L\indices{^\sigma_{\mu\nu}}-L\indices{^\lambda_{\sigma\mu}}L\indices{^\sigma_{\nu\lambda}})
\end{equation}
Here, $L\indices{^\lambda_{\sigma\mu}}$ is the disformation tensor which is defined as
\begin{equation}
\label{eqn:disformation tensor}
    L\indices{^\lambda_{\sigma\mu}}=-\frac{1}{2}g^{\lambda\gamma}(\nabla_\mu g_{\sigma\gamma}+\nabla_\sigma g_{\gamma\mu}-\nabla_\gamma g_{\sigma\mu})
\end{equation}
The non-metricity $Q$ can be calculated as $Q=-Q_{\lambda\mu\nu}P^{\lambda\mu\nu}$, where $Q_{\lambda\mu\nu}=\nabla_\lambda g_{\mu\nu}$ and $P\indices{^\lambda_{\mu\nu}}$ is the super-potential given by
\begin{equation}
    P\indices{^\lambda_{\mu\nu}}=\frac{1}{4}g_{\mu\nu}\left(Q^{\lambda}-\Tilde{Q}^{\lambda}\right)-\frac{1}{4}\delta^{\lambda}_{(\mu}Q_{\nu)}-\frac{1}{2}L\indices{^\lambda_{\mu\nu}}
\end{equation}
where $Q_\lambda=Q\indices{_\lambda^\mu_\mu}$ and $\Tilde{Q}_{\lambda}=Q\indices{^\mu_{\lambda\mu}}$ are the two traces. The energy-momentum tensor $T_{\mu\nu}$ is specified as
\begin{equation}
    T_{\mu\nu}=-\frac{2}{\sqrt{-g}}\frac{\delta(\sqrt{-g}\mathcal{L}_m)}{\delta g^{\mu\nu}}
\end{equation}
Varying Eqn. (\ref{eq:Action}) with respect to $g_{\mu\nu}$ gives the following modified Einstein's equation:

\begin{multline}
\label{eq:field_eqn}
    \frac{2}{\sqrt{-g}}\nabla_\lambda\left(\sqrt{-g}\left(f^{'}_1(Q)+2f^{'}_2(Q)\mathcal{L}_m\right) P\indices{^\lambda_{\mu\nu}}\right)+\frac{1}{2}g_{\mu\nu}f_1(Q)\\
    +\left(f^{'}_1(Q)+2f^{'}_2(Q)\mathcal{L}_m\right)\left(P_{\mu\lambda\sigma}Q\indices{_\nu^{\lambda\sigma}}-2Q_{\lambda\sigma\mu}P\indices{^{\lambda\sigma}_{\mu}}\right)
    =-f_2(Q)T_{\mu\nu}
\end{multline}
where $\{\hspace{2pt}^{'}\hspace{2pt}\}$ represents differentiation with respect to $Q$. In the next section, we find the field equations in the spatially flat FLRW universe. 
\section{Flat FLRW universe}
\label{sec:flatflrw}
Consider the flat FLRW metric
\begin{equation}
\label{eq:metric}
    ds^2=-dt^2+a^2(t)(dx^2+dy^2+dz^2)
\end{equation}
where $a(t)$ is the scale factor. In the backdrop of this metric, the non-metricity in the coincident gauge \cite{harko2018coupling}, using Eqs. (\ref{eqn:non-metricity}),(\ref{eqn:disformation tensor}) is given as 
\begin{equation}
    Q=6H^2
\end{equation}
where $H=\Dot{a}/a$ and $\{\hspace{2pt}{\Dot{}}\hspace{2pt}\}$ is the derivative with respect to time. We assume that the energy-momentum tensor $T_{\mu\nu}$ is given in the form of a perfect fluid 
\begin{equation}
\label{eq:T_mu_nu}
    T_{\mu\nu}=(p+\rho)u_{\mu}u_{\nu}+pg_{\mu\nu}
\end{equation}
where $\rho$ and $p$ are the energy density and pressure respectively. The four-velocity $u_{\mu}$ satisfies the normalization $u_{\mu}u^{\mu}=-1$. Substituting the metric Eq. (\ref{eq:metric}) and Eq. (\ref{eq:T_mu_nu}) in Eq. (\ref{eq:field_eqn}), we get the following Friedman equations:  
\begin{equation}
\label{eq:friedman_eqns}
    \begin{split}
        3H^2&=\rho_{\text{eff}}=\frac{f_2}{2f_Q}\left(-\rho+\frac{f_1}{2f_2}\right)\\
        \Dot{H}+3H^2&=-p_{\text{eff}}=\frac{f_2}{2f_Q}\left(p+\frac{f_1}{2f_2}\right)-\frac{\Dot{f}_Q}{f_Q}H
    \end{split}
\end{equation}
where $f_Q=f^{'}_1(Q)+2f^{'}_2(Q)\mathcal{L}_m$. Here $\rho_{\text{eff}}$ and $p_{\text{eff}}$ represent the effective energy density and pressure respectively. Using Eqs. (\ref{eq:friedman_eqns}), the continuity equation is given as  
\begin{equation}
    \Dot{\rho}+3H(p+\rho)=-6\frac{f^{'}_2}{f_2}H\Dot{H}(\mathcal{L}_m+\rho)
\end{equation}
If one sets $\mathcal{L}_m=-\rho$ \cite{harko2018coupling}, we retrieve the standard continuity equation:
\begin{equation}
\label{eq:continuity}
    \Dot{\rho}+3H(p+\rho)=0
\end{equation}
For the purposes of this work, we consider the universe to be comprised of baryons $(b)$ and GCG (denoted in the equations by $g$). We do not consider the interactions of GCG with baryons in this study. For pressureless baryonic matter, integrating Eq. (\ref{eq:continuity}) gives $\rho_{(b)}\propto a^{-1/3}$. We set the value of $a$ at the current epoch as 1 i.e. $a(t_0)=1$. Using $a(z)=1/(1+z)$, $\rho_{(b)}$ can be expressed as a function of redshift $z$
\begin{equation}
\label{eq:baryon_eos}
    \rho_{(b)}=\rho_{b,0}(1+z)^3
\end{equation}
where $\rho_{b,0}$ is the current value of baryonic energy density. The equation of state for the generalized Chaplygin gas \cite{bento2002generalized} is 
\begin{equation}
p_{(g)}=-\frac{A}{\rho_{(g)}^n}
\end{equation}
where $0\le n\le 1$ and $A$ are positive parameters. Using the continuity equation Eq. (\ref{eq:continuity}), the energy density $\rho_{(g)}$ as a function of scale factor $a$ is given as
\begin{equation}
    \rho_{(g)}(a)=\left(A+Ca^{-3(1+n)}\right)^{\frac{1}{1+n}}
\end{equation}
where $C$ is a constant of integration. One can re-write this expression as a function of $z$ by 
\begin{equation}
\label{eq:rho^g}
    \rho_{(g)}(z)=\left(A+C(1+z)^{3(1+n)}\right)^{\frac{1}{1+n}}
\end{equation}
Let $\rho_{(g)}(z=0)\equiv\rho_{g,0}$. One can  write Eq. (\ref{eq:rho^g}) as
\begin{equation}
\label{eq:chaplygin_eos}
    \rho_{(g)}(z)=\rho_{g,0}\left(A_s+(1-A_s))(1+z)^{3(1+n)}\right)^{\frac{1}{1+n}}
\end{equation}
where $A_s=A/\rho_{g,0}$ is the scaled parameter and $A_s>0$. Next, we introduce our model.  
\subsection{Model}
\label{sec:modelquad}
Motivated by quadratic corrections in $Q$ that were utilized in studies such as \cite{lu2019cosmology}, we assume the following functional forms for $f_1(Q)$ and $f_2(Q)$:
\begin{equation}
    f_1(Q)=\alpha Q+\beta Q^2, \hspace{10pt} f_2(Q)=Q.
\end{equation}
By making use of these two functional forms, and substituting Eqs.(\ref{eq:baryon_eos}),(\ref{eq:chaplygin_eos}) in Eq. (\ref{eq:friedman_eqns}), the expression for Hubble parameter becomes:
\begin{equation}
\label{eq:hubble_chaplygin}
    H(z)=H_0\sqrt{\left(\frac{1}{3\beta}\right)\left(\Omega_{g,0}\left(A_s+(1-A_s)(1+z)^{3(1+n)}\right)^{\frac{1}{1+n}}+\Omega_{b,0}(1+z)^3-\frac{\alpha}{6H_0^2}\right)}
\end{equation}
where $\Omega_{b,0}=\rho_{b,0}/(3H_0^2)$,  $\Omega_{g,0}=\rho_{g,0}/(3H_0^2)$ and $H(z=0)=H_0$. Using $H(z=0)=H_0$, we get the following relation
\begin{equation}
\label{eq:normalization}
    \Omega_{g,0}=-\Omega_{b,0}+3\beta+\frac{\alpha}{6H_0^2}
\end{equation}
We have a set of 5 unknown model parameters: $\{\alpha,\beta,A_s,\Omega_{b,0},n\}$. $\Omega_{g,0}$ can be constrained using Eq. (\ref{eq:normalization}). To obtain the values for the parameters, we discuss the datasets in the next section.  

\section{Datasets and Methods}
\label{sec:data}
We will be using two different datasets in order to constrain the aforementioned parameters; which will be introduced in the following. 
\subsection{H(z)}
\label{sec:hdata}
The Hubble parameter $H(z)$ in terms of redshift is given as 
\begin{equation}
    H(z)=-(1+z)\frac{dz}{dt}
\end{equation}
where $dz/dt$ is obtained in a model-independent way through spectroscopic dating of galaxy ages in the "differential age methods". Here we are working with Hubble measurements obtained from Baryonic Acoustic Oscillation (BAO), and the differential age methods that were found in studies \cite{hz1,hz2,hz3,hz4,hz5,hz6,hz7,hz8,hz9,hz10,hz11,hz12,hz13,hz14,hz15,hz16,hz17,hz18,hz19}. In \cite{Solanki_2021}, the authors have compiled a dataset of these $57$ points, which are comprised of the Hubble parameter $H(z)$, its associated error $\sigma_H$, and the corresponding redshift $z$.
We use the following $\chi^2$ function to constrain the parameters $\{\alpha,\beta,A_s,\Omega_{b,0},n\}$ using said dataset;
\begin{equation}
    \chi^2(\alpha,\beta,A_s,\Omega_{b,0},n)=\sum\limits_{i=1}^{57}\frac{(H_{\text{model}}(z_i,\alpha,\beta,A_s,\Omega_{b,0},n)-H_{\text{data}}(z_i))^2}{\sigma_{H(z_i)}^2}
\end{equation}
where $H_{\text{model}}(z_i)$ and $H_{\text{data}}(z_i)$ are the theoretical and observed values of the Hubble parameter at $z=z_i$ respectively. Based on \cite{Amati_2019}, we set $H_0=67.76$ $\mathrm{km} \mathrm{s}^{-1} \mathrm{Mpc}^{-1}$. For the Hubble data which is cosmology-independent, we use Equation \ref{eq:hubble_chaplygin} as $H_{\text{model}}(z_i)$ to obtain a fit with the aforementioned data.
We have made use of the Markov Chain Monte Carlo code EMCEE \cite{emcee2013} to constrain the unknown parameters; the results of which will be discussed in the section \ref{sec:results}.
\subsection{Quasars}
\label{sec:quasardata}
\cite{Lusso_2020} have constructed a table of $\approx 2400$ quasars, that contains the redshifts, and the X-ray and ultraviolet fluxes of these quasars, which we have taken from \cite{Lusso_2020}; from the VizieR catalog \cite{vizier}. By using the well-established X-ray-UV relation that quasars show \cite{Risaliti_2015}; one can deduce their distance moduli, which will be explained next.\\

\subsubsection{Cosmology-Independent Calibration}
\label{sec:circularity} 
One dilemma that arises when using these datasets for cosmological models other than $\Lambda$CDM is that the method used by \cite{Lusso_2020} to obtain the distance moduli assumes a $\Lambda$CDM model. This problem, which is also known as the circularity problem, can be avoided by constructing curvature-dependent luminosity distances, through the use of cosmic-chronometer measurements \cite{Wei_2020}. These are measurements obtained from passively evolving galaxies to measure $H(z)$ independent of any cosmological model \cite{2002ApJ...573...37J}. \cite{Wei_2020} have used this method in order to calibrate the distance moduli of the observations to other models. We will incorporate their method in this section since they have calibrated these values with a different dataset. 

In order to do so, we make use of the following fit  proposed by \cite{Amati_2019} for the Hubble constant $H(z)$ obtained through 31 cosmology-independent measurements, in order to calibrate the new version of the datasets that are used in \cite{Lusso_2020}:\\
\begin{equation}
    H_n(z)= \sum_{d=0}^n \beta_d h_n^d(z) 
    \label{eq:chronometer-fit}
\end{equation}
where $\beta_d$ are the coefficients of the Bernstein polynomial $h_n^d(z)$ given by:\\
\begin{equation}
    h_n^d(z) \equiv \frac{n!(z/z_m)^d}{d!(n-d)!}\left(1-\frac{z}{z_m}\right)^{n-d} 
    \label{eq:bernst}
\end{equation}
where $z_m$ is the maximum measured redshift in the 31 measurements which according to \cite{Wei_2020} is $z_m=2$.  \cite{Amati_2019} have set $n=2$, therefore the final form of the equation that we utilize is:\\
\begin{equation}
H(z) = \beta_0\left(1-\frac{z}{z_m}\right)^2 + 2\beta_1\left(1-\frac{z}{z_m}\right)\left(\frac{z}{z_m}\right) + \beta_2\left(\frac{z}{z_m}\right)^2 
\label{eq:hubble-nogeo}
\end{equation}
According to the fit obtained in \cite{Amati_2019}, $\beta_0=H_0=67.76\pm 3.68$, $\beta_1=103.33\pm 11.16$, and $\beta_2=208.45\pm 14.29$ in the units of $\mathrm{km} \mathrm{s}^{-1} \mathrm{Mpc}^{-1}$.

In order to be able to circumvent the circularity problem, we first start with the X-ray/UV relation \cite{Risaliti_2015}:

\begin{equation}
    \log{F_{x,\text{theory}}} =  \gamma \log{F_{UV}} + \beta ' +2(\gamma -1)\log(D_L^{\text{theory}}(z,\Omega_k))
    \label{eq:fxmlf}
\end{equation}
where $F_x$ and $F_{UV}$ are the X-ray and Ultraviolet fluxes respectively, $\beta^{'}$ and $\gamma$ are two unknown parameters that characterize the X-ray/UV relation, and $D_L^{\text{theory}}$ is the luminosity distance obtained from theory and is given by the following equations:
\begin{equation}
D_L^{\text{theory}}(\Omega_k,z)=
    \begin{cases}
    (1+z) \frac{c}{H_0 \sqrt{\Omega_k}}\sinh{\left[\frac{\sqrt{\Omega_k}D_C(z)}{D_H}\right] }, & \text{for } \Omega_k>0\\[10pt]
    (1+z) D_C(z) , & \text{for } \Omega_k=0 \\[10pt]
    (1+z) \frac{c}{H_0 \sqrt{\Omega_k}}\sin{\left[\frac{\sqrt{\lvert \Omega_k \lvert}D_C(z)}{D_H}\right] }, & \text{for } \Omega_k<0\\[10pt]
    \end{cases}
\end{equation}
where $\Omega_k$ is the curvature parameter and $D_C(z)$ is the co-moving distance and is obtained with the following equation:
\begin{equation}
    D_C(z)=c \int\limits_{0}^{z}\frac{dz^{'}}{H(z^{'})}
\end{equation}
\begin{figure}[ht!]
	\centering                            
	\includegraphics[width=15cm]{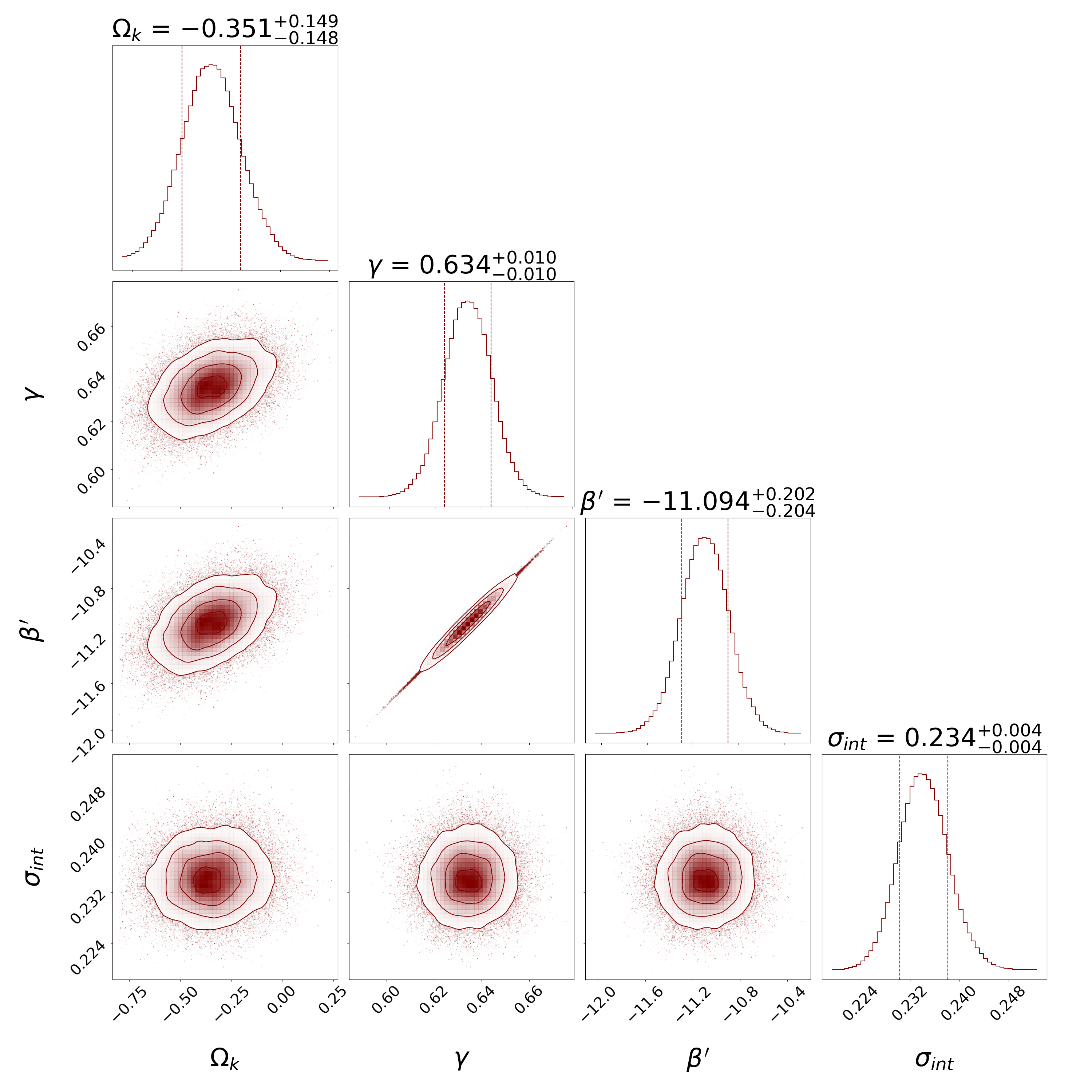}
	\caption{The corner plot above displays the two-dimensional contour plots for the 4 parameters $\{\Omega_k,\gamma,\beta^{'},\sigma_{\text{int}}\}$ with $1-\sigma$, $2-\sigma$ and $3-\sigma$ error bands that characterize the UV-X-ray relation, in a cosmology-independent way. This simulation was carried out with 20 walkers, for 5000 steps, 20$\%$ of which were burn-in steps.}
		\label{fig:circularity_corner}
\end{figure}
Then, the maximum likelihood function (MLF) can be constructed in the following way:\\
\begin{equation}
\log{\text{MLF}}=-0.5 \left\{\sum_i\left(\frac{\log{F_{x-\text{theory},i}}-\log{F_{x-\text{data},i}}}{\sigma_{tot,i}}\right)^2 + \log{2\pi \sigma_{\text{tot},i}^2}\right\}
\label{eq:logmlt}
\end{equation}
where
\begin{equation}
\sigma_{\text{tot},i}^2 = \sigma_{\text{int}}^2 + \left[2(\gamma-1)\frac{\sigma_{D_L^{\text{theory},i}}}{\ln{10D_L^{\text{theory},i}}}\right]^2,  
\label{eq:sigma}
\end{equation}
in which $\sigma_{\text{int}}$ is the internal dispersion to the tackle Eddington bias, and is another unknown parameter. Since we aren't assuming anything about the curvature parameter either; both $\Omega_K$ and $\sigma_{int}$ will also be assumed as free parameters in the calculations. Also, $\sigma_{D_L^{\text{theory},i}}$ is the uncertainty in luminosity distance; whose expressions we take from \cite{Wei_2020}.

In order to constrain the 4 parameters $\beta'$, $\gamma$, $\sigma_{int}$, and $\Omega_k$; we once again use Equation \ref{eq:hubble-nogeo} in order to obtain luminosity distances that we could input into Equation \ref{eq:fxmlf}. By carrying out this procedure and using what we have from the mentioned datasets, we could obtain a value for the 4 unknown parameters we have. Then we could use \ref{eq:fxmlf} with the obtained values for $\beta'$ and $\gamma$ to work with any other cosmological model; since they have been obtained without the assumption of any prior model.

One noteworthy matter is that since what exists in the catalogues isn't luminosity distance; rather the distance modulus $\mu$, we use the following standard relation to replace $D_L$ with $\mu$:
\begin{equation}
    \label{eq:mudl}
    \mu = 5\log{D_L^{theory}(z,\Omega_k)}+25
\end{equation}

In order to carry out this calibration, we have made use of the code EMCEE. We have adopted Gaussian priors for each of the parameters, and took a run with $20$ walkers and $5000$ steps, with $20$\% of them being counted as burn-in steps. The results can be seen in Fig. \ref{fig:circularity_corner}. Here, we observe that a mildly closed universe is preferred at a significance of $2.35\sigma$. This is different from the value obtained by the authors in \cite{Wei_2020} which stands at $2.14\sigma$. Therefore, in Sec. {\ref{sec:results}}, we have assumed zero spatial curvature $(\Omega_k=0)$ for our simulations. 

To test our model against the quasars dataset, we make use of the Eq. \ref{eq:fxmlf} with the parameter values obtained above. To calculate the errors for $\chi^2$, we use Gaussian error propagation. 
\section{Results}
In the following, we have set the variable $\Omega_{b,0}=0.05$, which has been determined in studies such as \cite{universe7100362}; leaving us with 4 parameters $\{\alpha,\beta,A_s,n\}$ to constrain.
\label{sec:results}
\subsection{H(z)}
\label{sec:hzresults}

\begin{figure}[ht!]
	\centering                            
	\includegraphics[width=17cm,height=19cm]{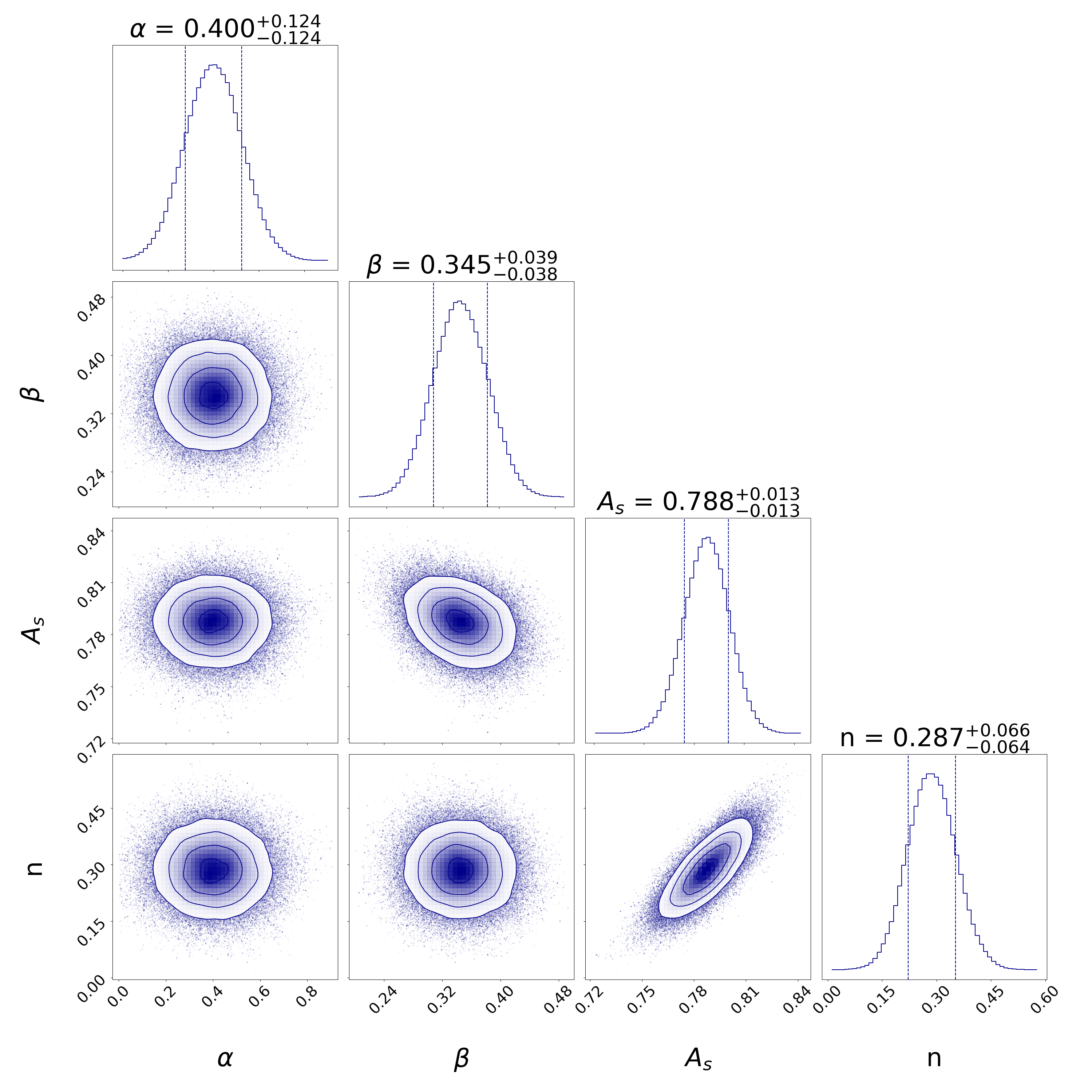}
	\caption{The corner plot above displays one-dimensional marginalized distributions and the two-dimensional contour plots for the 4 parameters $\{\alpha,\beta, A_s,n\}$ with $1-\sigma$, $2-\sigma$ and $3-\sigma$ error bands obtained with EMCEE using  $H(z)$ data. This was obtained with 30 walkers, and 10000 steps, 20$\%$ being burn-in steps.}
		\label{fig:corner_H}
\end{figure}

We have adopted Gaussian priors for each of the parameters, and set 30 walkers and 10000 steps, with $20 \%$ of them being counted as burn-in steps. The results of the simulation can be seen in the form of the corner plots in Fig. \ref{fig:corner_H}. Fig. \ref{fig:fit_results}-a shows the Hubble dataset plotted as a function of redshift, and compared with the values obtained from the corner plots; substituted in the fits.
\subsection{Quasars}
\label{sec:quasarresults}

\begin{figure}[ht!]
	\centering                            
	\includegraphics[width=17cm,height=19cm]{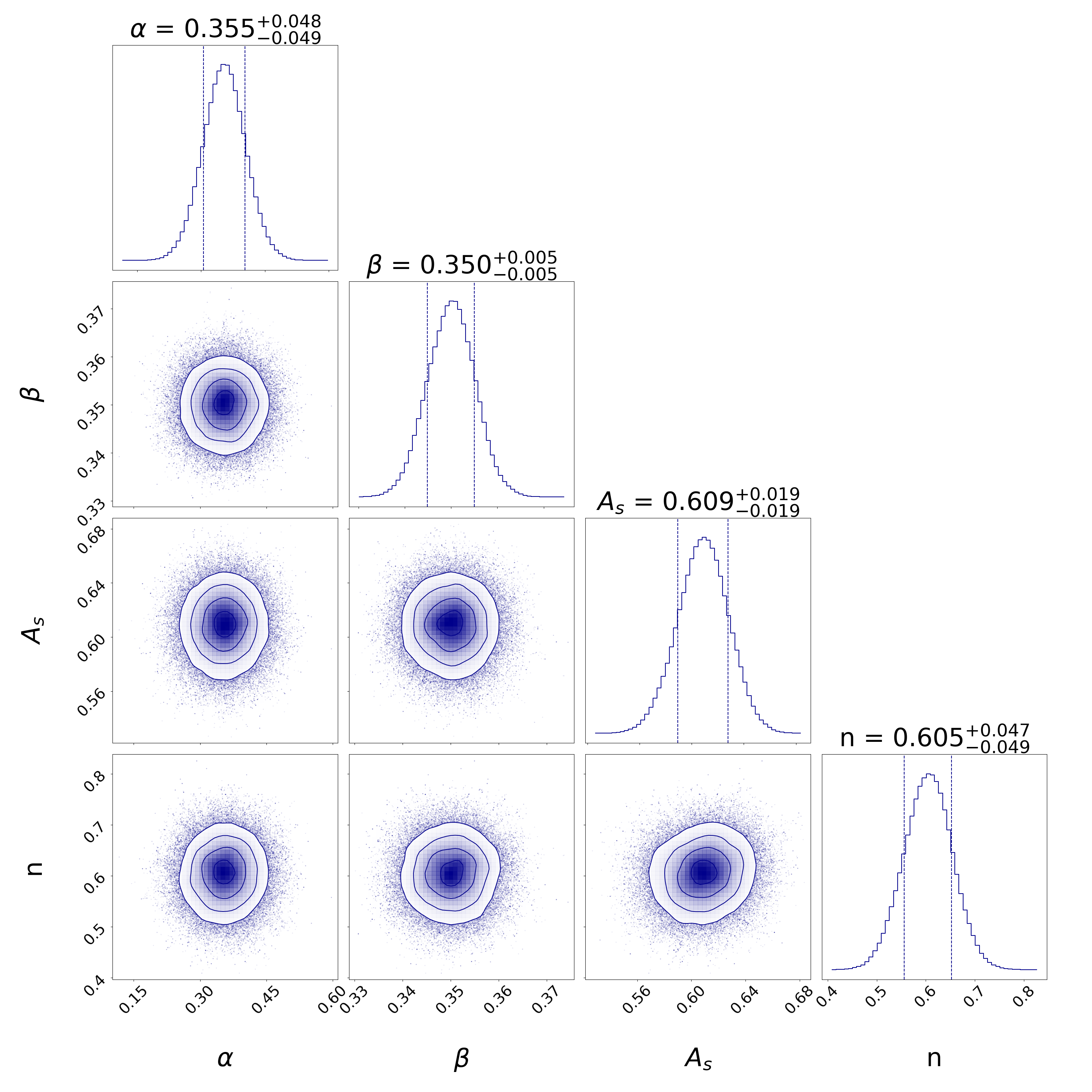}
	\caption{The corner plot above displays the one-dimensional marginalized distributions and two-dimensional contour plots for the 4 parameters $\{\alpha,\beta, A_s,n\}$ with $1-\sigma$, $2-\sigma$ and $3-\sigma$ error bands obtained with EMCEE using  quasar data from \cite{Lusso_2020}. This simulation was carried out with 20 walkers, for 10000 steps, 20$\%$ of which were burn-in steps.}
	\label{fig:corner_quasar}
\end{figure}

\begin{figure}[ht!]
    \centering
    \subfloat[\centering Hubble parameter $H$ against redshift $z$ for the Hubble dataset]{{\includegraphics[width=\textwidth, height=3.5in]{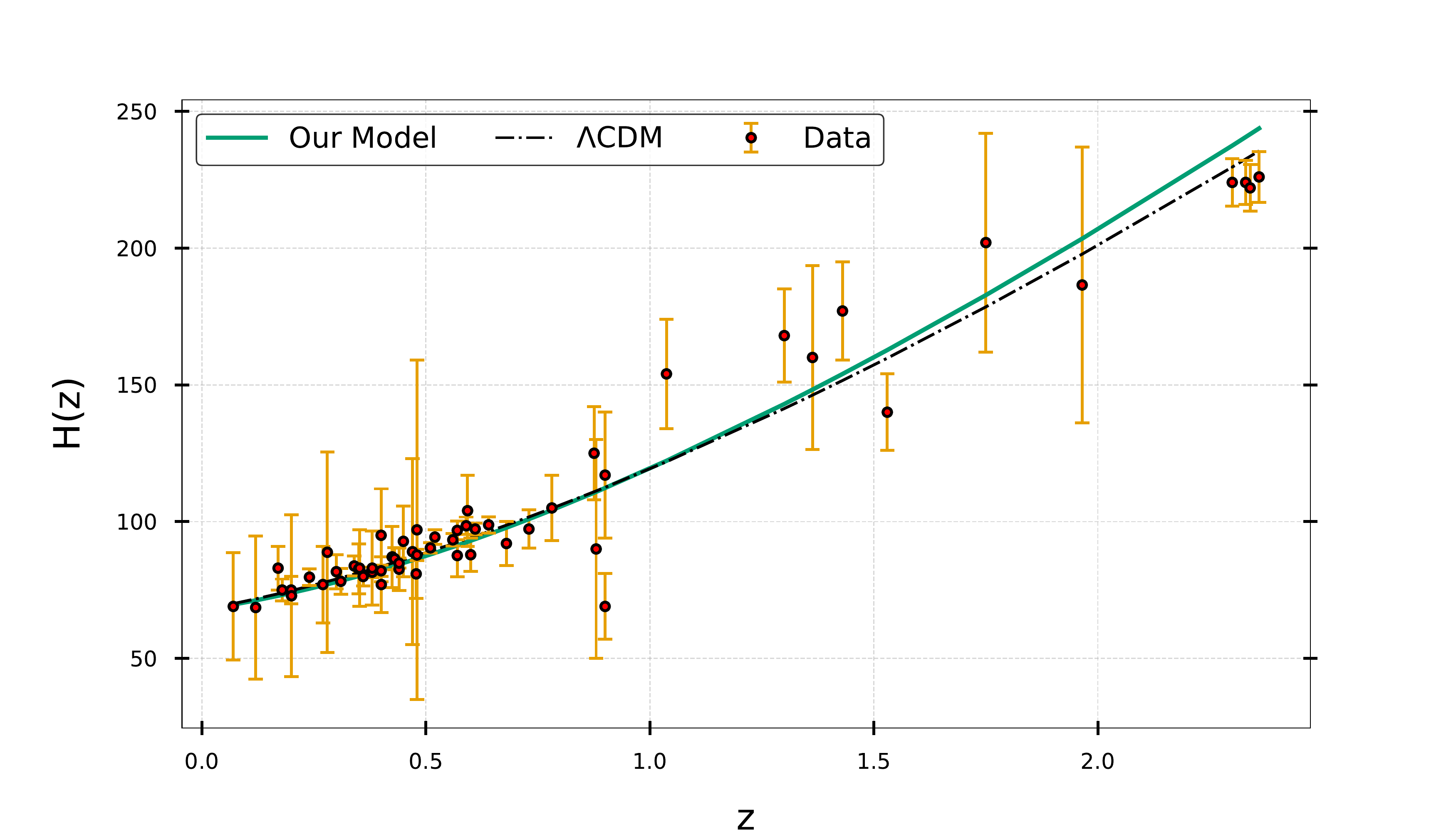}}}%
    \qquad\\
    \subfloat[\centering Distance Modulus $\mu$ against redshift $z$ for the quasar dataset]{{\includegraphics[width=\textwidth, height=3.5in]{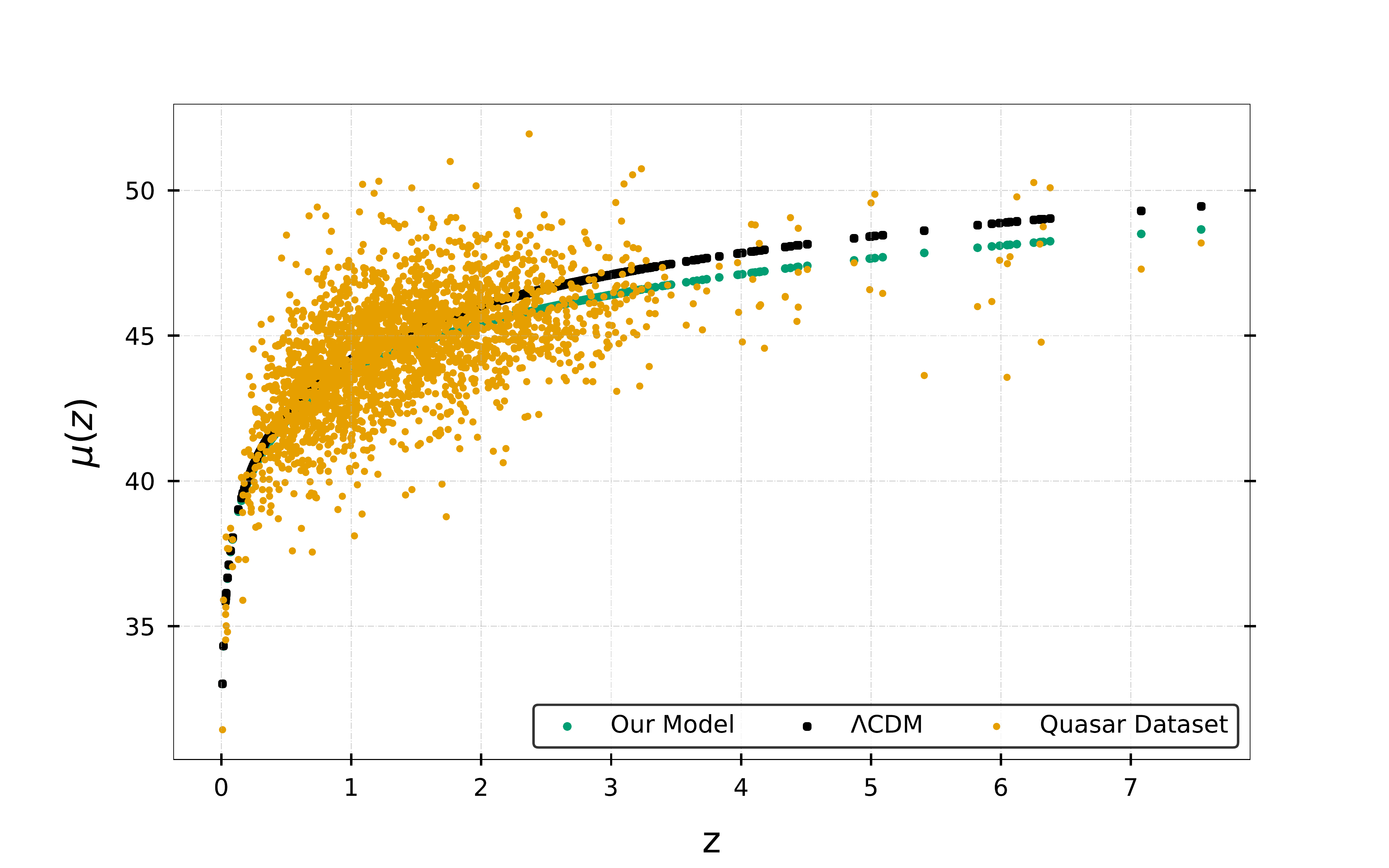} }}%
    \caption{(a) The solid line represents EMCEE fit of the proposed model to the $57$ datapoints of $H(z)$ dataset. $\Lambda$CDM model is indicated with the dashed-dot line. (b) The circular scatter marker represents the EMCEE fit of the proposed model to the $2421$ datapoints of quasar dataset. $\Lambda$CDM model is indicated with the square scatter marker. For the flat $\Lambda$CDM, we have assumed  the current value of cold dark matter density as $\Omega_m=0.3$.}%
    \label{fig:fit_results}%
\end{figure}

With the calibrated parameters obtained in \ref{sec:circularity}, we can now use the quasar dataset in order to constrain our model in the following manner. We make use of Eq. \ref{eq:fxmlf}; but this time we incorporate the $H(z)$  of equation \ref{eq:hubble_chaplygin} in order to calculate the luminosity distance.\\
We will once again use EMCEE \cite{emcee2013} in order to find the parameters $\{\alpha,\beta,A_s,n\}$. Here, we also assumed Gaussian priors; and took 20 walkers, and 10000 steps, with 20$\%$ of them being burn-in steps. The results can be seen in Fig. \ref{fig:corner_quasar}. 
In Fig. \ref{fig:fit_results}-b, we have plotted the distance modulus as a function of redshift, with the values obtained from the Monte-Carlo simulations; and have compared it with the data from \cite{Lusso_2020} and $\Lambda$CDM cosmology.
\subsection{Discussion}
\label{sec:disc}
The values of the parameters from H(z) and quasar datasets have been summarized in the table \ref{tab:constrained_both}. In our proposed model, the GCG plays the role of dark energy at late times. The corresponding energy density is characterized by parameters $\{A_s,n,\Omega_{g,0}\}$. To have at most luminal sound speed for perturbations \cite{bento2004revival}, $0<n\le 1$. For $n=1$, we get the case of ordinary Chaplygin gas (CG). The values we have obtained for $n$ do not support the CG but instead GCG. With both datasets, we get $0<A_s<1$, thereby indicating positive energy density. The positive value of $A_s$ was required for the stability of GCG perturbations  \cite{xu2012revisiting}. In  \cite{mandal2021constraint}, using the Pantheon dataset, the authors established that the power-law cosmology for a perfect fluid in non-minimally coupled $f(Q)$ gravity can give the $Q^2$ term within $2\sigma$, strongly suggesting that quadratic corrections are viable. In a recent paper \cite{thakur2017recent}, the values for parameters $A_s$ and $n$ were obtained as $0.722^{+0.021}_{-0.023}$ and $0.023^{+0.034}_{-0.034}$ respectively, by combining datasets from Observed Hubble Data (OHD), BAO, Cosmic Microwave Background (CMB) Shift and Supernovae (Union2.1). Their values for $A_s$ are within $2\sigma$ to our value obtained using $H(z)$ dataset. Their value for $n$ is in strong disagreement with our values. 

In \cite{chaplyginQT}, the authors also assumed a generalized Chaplygin gas in $f(Q,T)$ gravity where $T$ is the trace of the energy-momentum tensor. For their proposed Model-II in the paper, they found $n=0.53^{+0.14}_{-0.12}$. This value is within $2\sigma$ and $1\sigma$ in comparison to our model with the H(z) and quasar datasets respectively, thus in good agreement. In the subsequent section, we study the kinematic diagnostics to test our model.
\FloatBarrier
\FloatBarrier
\hspace{-19.5pt} 
\bgroup
\def\arraystretch{1.5}
\begin{table}[ht!]
    \centering
    \begin{tabular}{|c|c|c|c|c|c|c|c|}
    \hline
    Dataset & $\alpha$ & $\beta$ & $A_s$ & $n$ & $\Omega_{g,0}$ & \text{dof} &$\chi^2/\text{dof}$\\
    \hline 
    H(z)    & $0.400^{+0.124}_{-0.124}$ &$0.345^{+0.039}_{-0.038}$ & $0.788^{+0.013}_{-0.013}$& $0.287^{+0.066}_{-0.064}$&0.985&56& 1.220\\ 
    \hline 
    Quasars    & $0.355^{+0.048}_{-0.049}$  &$0.350^{+0.005}_{-0.005}$ & $0.609^{+0.019}_{-0.019}$& $0.605^{+0.047}_{-0.049}$&1.000&2420&0.674\\
    \hline
    \end{tabular}
    \caption{Summary of MCMC results for the model's parameters for the two datasets. Here, dof represents the degrees of freedom.}
    \label{tab:constrained_both}
\end{table}
\section{Diagnostics}
\label{sec:diagnostics}
We now investigate whether our proposed model can explain the accelerated expansion of the universe by studying kinematic diagnostics like deceleration and jerk parameters. 
\subsection{Deceleration Parameter}
\label{sec:q}

The deceleration parameter $q$ indicates whether the universe undergoes an accelerated expansion or not. If $q<0$, it indicates an accelerated expansion. An initial decelerating phase is required for the process of structure formation; whereas an accelerating phase in late time can explain the current observations of an accelerating expansion. 
This indicates that a phase transition $q=0$ must occur at transition redshift $z_t$. Deceleration $q$ can be defined in terms of the Hubble Parameter as
\begin{equation}
    q(z)=-1+\frac{(1+z)}{H(z)}\frac{dH(z)}{dz}
\end{equation}
For our model, $q(z)$ can be expressed as
\begin{equation}
    q(z)=-1+\frac{3}{2}(1+z)^3\frac{\phi_1}{\phi_2}
\end{equation}
where $\phi_1$ and $\phi_2$ are respectively given as
\begin{equation}
   \begin{split}
    \phi_1&=\Omega_{b,0}+\Omega_{g,0}(1-A_s)(1+z)^{3n}\left(A_s+(1-A_s)(1+z)^{3(1+n)}\right)^{-\frac{n}{1+n}}\\
    \phi_2 &= \left(\Omega_{g,0}\left(A_s+(1-A_s)(1+z)^{3(1+n)}\right)^{\frac{1}{1+n}}+\Omega_{b,0}(1+z)^3-\frac{\alpha}{6H_0^2}\right)
\end{split} 
\end{equation}
Let $q_0$ denote the current value of $q$. These values are listed in the table \ref{tab:diagnostics_values}. For our suggested model, $z_t$ constrained through the $H(z)$ dataset is close to the $z_t$ for $\Lambda$CDM as compared to the model constrained through quasar datasets.    
\begin{table}[h]
    \centering
    \begin{tabular}{|c|c|c|c|}
    \hline 
         & $q_0$ & $z_t$ & $j_0$  \\
    \hline 
    H(z)     & $-0.625$ & $0.607$ & 1.205\\
    \hline 
    Quasars     & $-0.370$ & $0.240$ & 1.617 \\
    \hline 
    $\Lambda$CDM     & $-0.550$ & $0.671$ & 1.000\\
    \hline 
    \end{tabular}
    \caption{The values of the current deceleration $q_0$, the transition redshift $z_t$ in the deceleration parameter and the current jerk parameter $j_0$ are listed in this table.}
    \label{tab:diagnostics_values}
\end{table} 
\subsection{Jerk Parameter}
\label{sec:j}

The jerk parameter $j$ is the fourth term in the Taylor series expansion of the scale factor about its present value. It is another kinematic diagnostic that measures deviations from the $\Lambda$CDM model. One can write the jerk parameter $j$ in terms of the deceleration parameter $q$ as
\begin{equation}
    j(z)=q(z)+2q(z)^2+(1+z)\Dot{q}(z)
\end{equation}
Its value for $\Lambda$CDM universe is 1 and is independent of the redshift. For our model, $j(z)$ is
\begin{equation}
    \begin{split}
        j(z)&=1+\frac{3}{2}(1+z)^4\frac{\phi_3}{\phi_2}
    \end{split}
\end{equation}
where $\phi_3$ is given as
\begin{equation}
\begin{split}
    \phi_3&=3n\Omega_{g,0}(1-A_s)(1+z)^{3n-1}\left(A_s+(1-A_s)(1+z)^{3(1+n)}\right)^{-\frac{n}{1+n}}\\
    &\hspace{14pt}-3n\Omega_{g,0}(1-A_s)^2(1+z)^{2(3n+1)}\left(A_s+(1-A_s)(1+z)^{3(1+n)}\right)^{-\frac{1+2n}{1+n}}
\end{split}
\end{equation}
Let $j_0$ be the current value of jerk. The $j_0$ values for the 2 datasets are listed in table \ref{tab:diagnostics_values}. From fig. (\ref{fig:diagnostics}-b), we observe that initially $j$ is close to 1 and decreases with increasing redshift after $z\approx0.5$ for both datasets. 
\begin{figure}%
    \centering
    \subfloat[\centering Deceleration Parameter]{{\includegraphics[width=6.5cm,height=6.25cm]{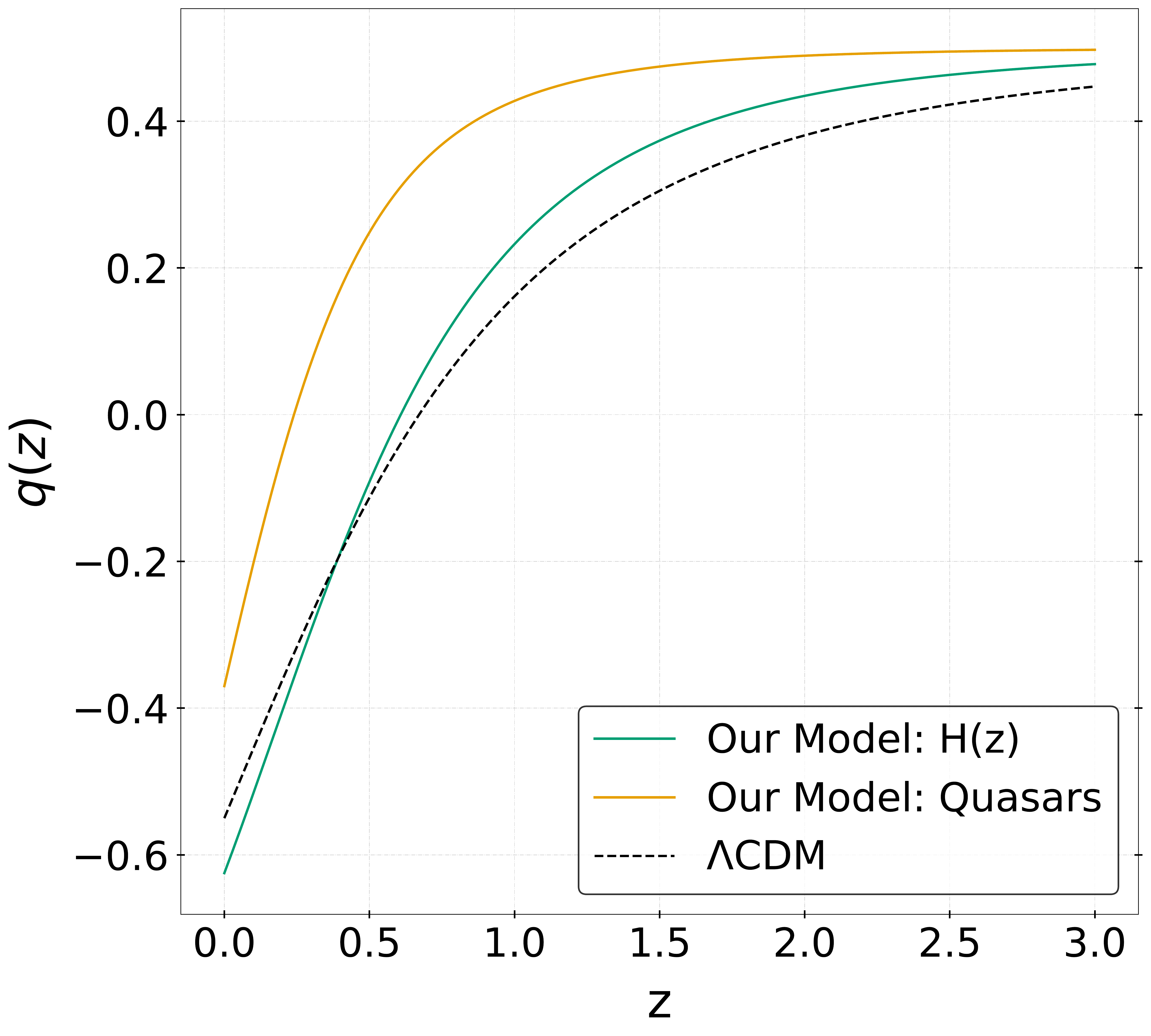} }}%
    \qquad
    \subfloat[\centering Jerk Parameter]{{\includegraphics[width=7cm,height=7cm]{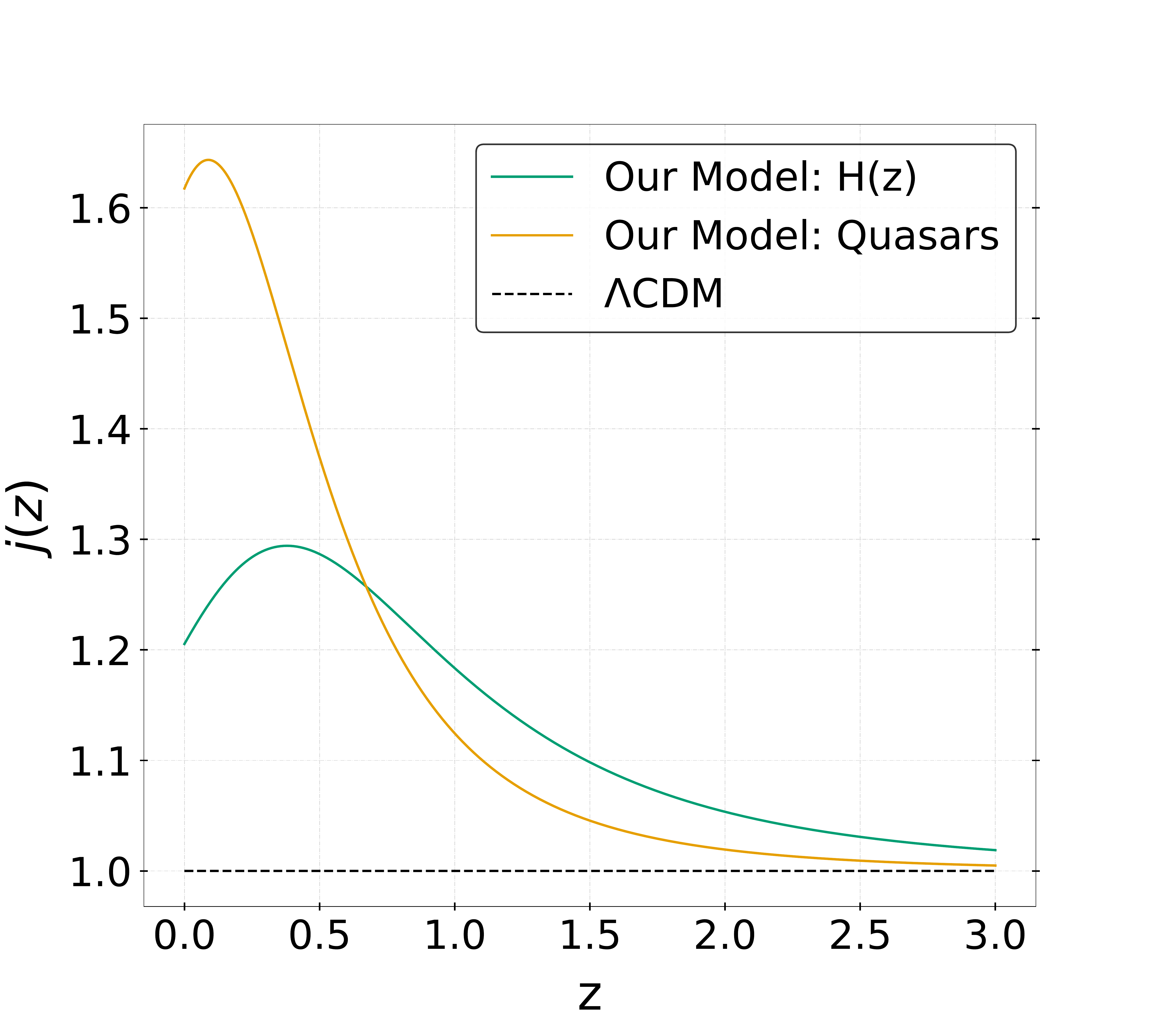} }}%
    \caption{Deceleration and Jerk parameter plotted with the two datasets. For the flat $\Lambda$CDM, we have assumed  the current value of cold dark matter density as $\Omega_m=0.3$.}%
    \label{fig:diagnostics}%
\end{figure}
\section{Conclusion}
\label{sec:conclusions}
In this work, we studied the $f(Q)$ cosmology non-minimally coupled to matter. We introduced a quadratic correction term to the standard $f(Q)=Q$ theory such that in our suggested model, we considered the functional form $\alpha Q+\beta Q^2$ and took a linear $Q$ coupling with the matter. We assumed a pressureless baryonic matter and GCG as the background fluid. The GCG is characterized by parameters $n$ and $A_s$. For $n=0$, one gets the standard model of cosmology, $\Lambda$CDM. We used Markov Chain Monte Carlo (MCMC) methods to constrain our 4-parameter space $\{\alpha,\beta, A_s,n\}$ using $57$ $H(z)$ measurements from BAO and differential age methods and $2421$ measurements from quasars. We obtained $n=0.287^{+0.066}_{-0.064}$ and $n=0.605^{+0.047}_{-0.049}$ with the two datasets respectively, highlighting that our model deviates from $\Lambda$CDM. Additionally, our results for $n$ are in good agreement with the Model-II proposed by authors in \cite{chaplyginQT}. 

The obtained values of $\alpha$ and $\beta$ for the two datasets fall within $1\sigma$ of each other. The same however cannot be said for $A_s$. In \cite{li2018constraint}, the authors used the Planck 2015 Cosmic Microwave Background Anisotropy, type-Ia Supernovae, Observed Hubble Data sets to constrain GCG. They found that $A_s=0.759^{+0.020}_{-0.032}$. This value is in good agreement ($2\sigma$) with the value we got from $H(z)$ dataset. But, their analysis predicts a resemblance of the behavior of adiabatically perturbed GCG to $\Lambda$CDM which is in sharp contrast to our analysis. Nonetheless, for both $A_s$ and $n$, the values we obtained are in the expected ranges: $0< n\le 1$; $A_s>0$.  

Also noteworthy is the fact that through analysis of the deceleration parameter $q$, we found that with both datasets, there is evidence for an accelerated expansion of the universe, in accordance with numerous studies such as \cite{riess1998observational,perlmutter1999measurements,eisenstein2005detection,percival2010baryon}. The transition redshift $z_t$ at which the universe switches from a decelerated to an accelerated phase was obtained as $0.607$ and $0.240$ for the two datasets with the current values of $q$ as $-0.625$ and $-0.370$ respectively. The results from $H(z)$ dataset are consistent with numerous studies \cite{magana2014cosmic,mamon2017parametric,farooq2017hubble}. However, $z_t$ result from the quasar data are less compatible with the existing literature. We also studied the jerk parameter $j$ to validate our model. The jerk $j_0$ of our proposed model at the current epoch shows deviations from the $\Lambda$CDM universe.   

Finally, it should be mentioned that the results of our studies could be improved if more measurements of cosmic chronometers become available. The fit we used to calibrate the data taken from \cite{Wei_2020}, predicts a mildly closed universe at $2.35\sigma$. This is in fact a result that \cite{Wei_2020} also obtained as well at $2.14\sigma$; albeit with a different older version of the quasar dataset which comprises of lesser number of data points. 
Since there are only 31 cosmic-chronometer measurements available up to $z=2$, we predict a more accurate fit would play a significant role in making the calibrations more precise and setting tighter constraints on the parameters. Furthermore, we know that $f(Q)$ cosmology does not suffer from the cosmological constant problem and explains the current observations well. This sets the motivation to test various other dark energy candidates such as modified models of Chaplygin gas in the light of non-minimal coupling. This can be a dedicated future research topic.  
\acknowledgments
F. S is grateful to the University of Tehran for supporting this work under a grant provided by the university research council.

This research has made use of the VizieR catalog access tool, CDS,
 Strasbourg, France (DOI : 10.26093/cds/vizier). The original description of the VizieR service was published in 2000, A\&AS 143, 23.
\bibliographystyle{jcap}

\bibliography{refs}

\end{document}